\documentclass[12pt]{article}

\usepackage{graphicx}
\usepackage{amsmath}
\usepackage{amssymb}

\def\eq#1{(\ref{#1})}
\def\[#1\]{\begin{align}#1\end{align}}
\begin{document}
\begin{titlepage}
\title{
\hfill\parbox{4cm}{ \normalsize YITP-13-11}\\
\vspace{1cm} A canonical rank-three tensor model \\with a scaling constraint}
\author{Naoki {\sc Sasakura}
\thanks{\tt sasakura@yukawa.kyoto-u.ac.jp}
\\[15pt]
{\it Yukawa Institute for Theoretical Physics, Kyoto University,}\\
{\it Kyoto 606-8502, Japan}}
\date{}
\maketitle
\thispagestyle{empty}
\begin{abstract}
\normalsize
A rank-three tensor model in canonical formalism has recently been proposed.
The model describes consistent local-time evolutions of fuzzy spaces through a 
set of first-class constraints which form an on-shell closed algebra with structure
functions. In fact, the algebra provides an algebraically consistent discretization of
the Dirac-DeWitt constraint algebra in the canonical formalism of general relativity.  
However, the configuration space of this model contains obvious degeneracies of
representing identical fuzzy spaces.
In this paper,
to delete the degeneracies, another first-class constraint representing a scaling symmetry 
is added to propose a new canonical rank-three tensor model. 
A consequence is that, 
while classical solutions of the previous model have typically 
runaway or vanishing behaviors,
the new model has a compact configuration space and its
classical solutions asymptotically approach either fixed points or cyclic orbits in time evolution. 
Among others, fixed points contain configurations with group symmetries, and
may represent stationary symmetric fuzzy spaces. 
Another consequence on the uniqueness of the local Hamiltonian
constraint is also discussed, and a minimal canonical tensor model,
which is unique, is given. 
\end{abstract}
\end{titlepage}

\section{Introduction}
\label{sec:introduction}
The tensor models have first been proposed
\cite{Ambjorn:1990ge,Sasakura:1990fs,Godfrey:1990dt}
as analytical description of the 
$D>2$ dimensional simplicial quantum gravity with hope to extend the success of
the matrix models for the $D=2$ dimensional case to the other dimensions.
The idea of the tensor models has also been applied to  
the loop quantum gravity as group field theories
by considering group-valued indices 
\cite{Boulatov:1992vp,Ooguri:1992eb,DePietri:1999bx,Freidel:2005qe,Oriti:2011jm}.
In these approaches, the theoretical interpretation of the tensor models 
is essentially based on the correspondence between perturbative 
Feynman diagrams of the tensor models and 
the dual diagrams of simplicial manifolds.
In the original tensor models with Hermitian tensors, however, 
the correspondence has delicate issues \cite{Sasakura:1990fs,DePietri:2000ii},  
and it is not known how to take
the large $N$ limit, which was essential in relating the matrix models 
to $D=2$ quantum gravity.   
On the other hand, another kind of tensor models with unsymmetric tensors,
called colored tensor models \cite{Gurau:2009tw}, have been proposed.
The colored tensor models have good correspondence to simplicial manifolds, 
and various analytical results including the large $N$ limit 
have been revealed \cite{Gurau:2011xp}. 
The colored tensor models have
also stimulated developments of renormalization of the tensor group field 
theories \cite{BenGeloun:2011rc,BenGeloun:2012pu,Geloun:2012bz,Carrozza:2012uv}.
However, the present situation of the tensor models as quantum gravity is still uncertain;
in Feynman perturbation series, the large $N$ limit of the colored tensor models 
is dominated by the ``melonic" diagrams
\cite{Gurau:2011xp,Bonzom:2012wa},
which are topologically spheres but look rather singular \cite{Gurau:2013cbh} unlike our actual space.
The dominance of the melonic diagrams in the large $N$ limit has also been shown \cite{Dartois:2013he} 
for other new models which are called multi-orientable tensor models \cite{Tanasa:2011ur}. 
  
In view of the present unsatisfactory status of the tensor models as quantum gravity 
in the above interpretation,
it would also be meaningful to pursue another interpretation of the tensor models. 
In fact, the present author has proposed the interpretation
that the rank-three tensor models, which have a rank-three tensor as their only
dynamical variable, may be regarded as dynamical models of fuzzy spaces
\cite{Sasakura:2011ma,Sasakura:2005js}.
An advantage of this interpretation is that, since 
fuzzy spaces can generally describe any dimensional spaces,
any dimensional quantum gravity can be considered to be incorporated in
the rank-three tensor models. 
This is in contrast with that ranks of tensors are directly related to dimensions
in the above interpretation in terms of simplicial manifolds.
In fact, by semi-classical analyses, 
the present author has shown spontaneous generation of various dimensional
fuzzy spaces \cite{Sasakura:2006pq} and Euclidean general relativity on them from a certain fine-tuned
rank-three tensor model \cite{Sasakura:2008pe,Sasakura:2009hs}.

However, the above results of the Euclidean tensor model are not satisfactory.
The action is complicated and unnatural. Moreover it must be fine-tuned 
so that the above physically wanted results be obtained, 
but there is no principle to choose the action
out of the other infinitely many possibilities. 
This drawback may be solved by a kind of universality through quantum mechanical 
treatment. But first of all it is necessary to introduce a notion of time into tensor models 
before discussing quantum mechanics.

Thus, to incorporate time into tensor models, the present author 
has proposed a rank-three tensor model in a canonical formalism
\cite{Sasakura:2012fb,Sasakura:2011sq}.
The model is defined as a pure constraint system with 
a set of first-class constraints which form an on-shell closed algebra
with structure functions. 
In fact, the algebra has a resemblance to
the Dirac-DeWitt first-class constraint algebra in the canonical formalism of general relativity
\cite{Arnowitt:1962hi,DeWitt:1967yk,Hojman:1976vp},
and the former agrees with the latter in a formal limit of vanishing fuzziness.
Moreover, there exist a notion of local time and local time evolutions controlled 
by local Hamiltonian constraints in the model, as in general relativity.
The on-shell closure condition is so strong that 
the local Hamiltonian constraints are (two-fold) unique under 
some physically reasonable assumptions.

However, as will be discussed below, the canonical
rank-three tensor model above seems to have 
some unsatisfactory features concerning the classical solutions. 
So the main purpose of the present paper is to propose
a new canonical rank-three tensor model by adding a constraint 
representing a scaling symmetry to the previous model.
The scaling symmetry is natural from the perspective of fuzzy spaces,  
and the new model has nice features for future study.  
The paper is organized as follows. In Section~\ref{sec:previous}, 
the canonical rank-three tensor model proposed in the previous paper
\cite{Sasakura:2012fb} is summarized.
In Section~\ref{sec:new}, 
the unsatisfactory features of the previous model are discussed, and 
a new model is proposed by adding   
a new first-class constraint representing a scaling symmetry.
In Section~\ref{sec:classical}, the configuration space and 
fixed points of the classical equation of motion of the new model are discussed.
Among others, such fixed points contain configurations with group symmetries.
In Section~\ref{sec:uniqueness}, the uniqueness of the local Hamiltonian constraint 
for the tensor model with a totally symmetric rank-three tensor is discussed. 
This provides a minimal canonical tensor model.
The finial section is devoted to summary and future prospects.

\section{The previous canonical rank-three tensor model} 
\label{sec:previous}
In this subsection, I will summarize the canonical rank-three tensor model proposed
in the previous paper \cite{Sasakura:2012fb}.

The dynamical variables of the canonical rank-three tensor model are 
given by the canonical variables, 
$M_{abc},\ P_{abc}\ (a,b,c=1,2,\cdots,N)$.
They satisfy the generalized Hermiticity condition,
\begin{align}
\label{eq:hermiticity}
X_{abc}=X_{bca}=X_{cab}=X^*_{bac}=X^*_{acb}=X^*_{cba},
\end{align}
where $X=M,P$ and $^*$ denotes complex conjugation.
The Poisson brackets between them are given by 
\begin{align}
\label{eq:funpoi}
&\{ M_{abc} ,P_{def} \}=\delta_{ad}\delta_{be}\delta_{cf}
+\delta_{ae}\delta_{bf}\delta_{cd} +\delta_{af}\delta_{bd}\delta_{ce},\\
&\{ M_{abc} ,M_{def} \}=\{ P_{abc} ,P_{def} \}=0.
\end{align}
Here the first Poisson bracket is taken to be consistent with the generalized Hermiticity condition \eq{eq:hermiticity}.

The kinematical symmetry of the canonical tensor model 
is given by the orthogonal group $O(N)$,
\begin{align}
\label{eq:kinematical}
X_{abc}=G_{ad}G_{be}G_{cf} X_{def}, \ \ G\in O(N),
\end{align}
where repeated indices are summed over. In what follows, this convention
is used, unless otherwise stated.

With the canonical variables, the Lie generators of the kinematical symmetry are 
expressed by
\[
\label{eq:defofcald}
{\cal J}_{[ab]}=\frac{\sigma}{2} \left( X_{acd}Y_{bcd}-X_{bcd}Y_{acd} \right ),
\]
where the square bracket $[\ ]$ in the index symbolically represents 
the antisymmetry, ${\cal J}_{[ab]}=-{\cal J}_{[ba]}$.  
As for $X,Y$, the following two cases, 
\[
\label{eq:xmyp}
\hbox{(i) }&X=M, \ Y=P, \\
\label{eq:xpym}
\hbox{(ii) }&X=P, \ Y=M,
\]
can be considered.
The numerical factor $\sigma$ in \eq{eq:defofcald} takes
for convenience the values,
\[
\label{eq:defofsigma}
\sigma=
\left\{
\begin{array}{cl}
-1 &\hbox{ for (i)},\\
1 & \hbox{ for (ii)},
\end{array}
\right.
\]
respectively.
With \eq{eq:defofsigma}, the fundamental Poisson bracket \eq{eq:funpoi} can 
be expressed as
\[
\{ X_{abc},Y_{def} \}=-\sigma (\delta_{ad}\delta_{be}\delta_{cf}
+\delta_{ae}\delta_{bf}\delta_{cd} +\delta_{af}\delta_{bd}\delta_{ce})
\] 
for both cases (i) and (ii).

The two consistent local Hamiltonian constraints, which
have a slight difference in index contraction, are given 
by\footnote{Strictly speaking, the previous paper \cite{Sasakura:2012fb} 
only deals with the case (ii).
As for the case (i),  ${\cal H}_a$ satisfies the conditions of the previous paper, 
if the time reversal symmetry is replaced with ${\cal H}_a\rightarrow -{\cal H}_a$.}
\[
\label{eq:hycde}
{\cal H}_a=X_{a(bc)}X_{bde}Y_{cde}, \\
\label{eq:hyced}
{\cal H}_a=X_{a(bc)}X_{bde}Y_{ced},
\]
where $X_{a(bc)}=(X_{abc}+X_{acb})/2$.

${\cal H}_a$ and ${\cal J}_{[ab]}$ form a Poisson algebra given by
\[
\label{eq:alghh}
&\{ H(T_1),H(T_2) \}=J([\tilde T_1,\tilde T_2]), \\
\label{eq:algjh}
&\{ J(V),H(T) \}=H(VT),\\
\label{eq:algjj}
&\{ J(V_1),J(V_2) \}=J([V_1,V_2]),
\]
where
\[
&H(T)=T_a {\cal H}_a, \\
&J(V)=V_{[ab]} {\cal J}_{[ab]},
\]
with a real vector $T_a$ and an antisymmetric real matrix $V_{[ab]}=-V_{[ba]}$.
On the right-hand sides of the Poisson algebra,
\[
\tilde T_{(bc)} = T_a X_{a(bc)},
\]
$VT$ is the usual multiplication of a matrix and a vector, and $[\ ,\ ]$ denotes 
the matrix commutator.  
Since the right-hand side of \eq{eq:alghh} contains $\tilde T$
dependent on $X$, the algebra has structure functions, but not
structure constants. 
This feature makes the apparently simple Poisson algebra
\eq{eq:alghh}, \eq{eq:algjh}, \eq{eq:algjj} highly non-trivial, and 
plays an essential role 
in deriving from the Poisson algebra
the Dirac-DeWitt first-class constraint algebra in
the canonical formalism of general relativity \cite{Arnowitt:1962hi,DeWitt:1967yk,
Hojman:1976vp}
by taking a formal limit of vanishing fuzziness \cite{Sasakura:2011sq}. 
It is also an important fact that the multiple possibilities \eq{eq:xmyp}, \eq{eq:xpym}, \eq{eq:hycde},
\eq{eq:hyced} actually lead 
to the same Poisson algebra \eq{eq:alghh}, \eq{eq:algjh}, \eq{eq:algjj}.

The closure of the Poisson algebra \eq{eq:alghh}, \eq{eq:algjh}, \eq{eq:algjj}
on the on-shell subspace defined by ${\cal J}_{[ab]}={\cal H}_a=0$ 
implies that a canonical
rank-three tensor model can consistently be defined as a constraint system
with a set of first-class constraints, 
 ${\cal J}_{[ab]}={\cal H}_{a}=0$.
 In analogy with general relativity, ${\cal J}_{[ab]}$ and ${\cal H}_a$ may
 be called the momentum and Hamiltonian constraints, respectively.

\section{A new canonical rank-three tensor model}
\label{sec:new}
As explained in Section~\ref{sec:previous},
the canonical tensor model is a pure constraint system with the first-class constraints,
 ${\cal J}_{[ab]}={\cal H}_{a}=0$.
Following the standard method for singular systems,
the total hamiltonian is given by
\[
H_{tot}={\cal N}_a {\cal H}_a+{\cal N}_{[ab]} {\cal J}_{[ab]},
\]
where
${\cal N}_a,\ {\cal N}_{[ab]}$ are arbitrary variables, the actual values of which may be 
fixed by some gauge fixing conditions.
For the choice of the local Hamiltonian \eq{eq:hycde}, 
the classical equation of motion for $X$ is given by
\[
\label{eq:ceqm}
\frac{d X_{abc}}{dt}=\{X_{abc},H_{tot}\}\approx 
-\sigma {\cal N}_d 
\left(X_{d(ae)}X_{ebc}+X_{d(be)}X_{eca}+X_{d(ce)}X_{eab}
\right)+{\cal N}_{[de]}(\cdots),
\]
where $\approx$ denotes the so-called weak equality, and $\cdots$ are
the terms representing 
the infinitesimal $O(N)$ transformation. 
The choice \eq{eq:hyced} as ${\cal H}_a$ instead of \eq{eq:hycde} will change 
the order of $abc$ on the right-hand side of \eq{eq:ceqm}, 
but this is not important for the following discussions. 

It is not difficult to numerically study the equation of motion \eq{eq:ceqm} 
simultaneously taking 
into account the constraints ${\cal H}_a={\cal J}_{[ab]}=0$ and 
some appropriate gauge fixing conditions.
This has been carried out, and it has turned out that 
the time-dependence of the classical solutions is rather extreme.  
This can essentially be captured 
by considering the following simplified version of \eq{eq:ceqm},
\[
\frac{dx}{dt}=x^2.
\]
The behavior in time evolution is obviously given by
\[
\begin{array}{ll}
x(t)\rightarrow \infty&\hbox{for }x(0)>0, \\
x(t)=0&\hbox{for } x(0) = 0, \\
x(t)\rightarrow -\,0 & \hbox{for }x(0)<0,
\end{array}
\]
for initial values $x(0)$.
Thus the point $x=0$ is the only fixed point, and 
$x(t)$ either diverges or asymptotically vanishes for non-vanishing initial values. 
From the numerical study, it seems that
the original equation \eq{eq:ceqm} has similar properties. 
There does not seem to exist any other fixed points but the trivial one $^\forall X_{abc}=0$, 
and $X_{abc}$ seem to either 
diverge or asymptotically vanish for non-trivial initial configurations.
These extreme behaviors cast doubts on the physical sense of the model.

On the other hand, in the numerical study, it has often been observed that
the ratios $X_{abc}/X_{def}$ have finite and non-vanishing asymptotic values.
This suggests that the model should be modified so that
only the ratios of $X_{abc}$ become the true dynamical variables.
This can easily be realized by introducing a gauge symmetry of 
common rescaling,
\[
\label{eq:gammax}
X_{abc}\rightarrow \gamma X_{abc} \ \hbox{   for all }X_{abc}, 
\] 
where $\gamma$ is real and arbitrary.

The gauge symmetry \eq{eq:gammax} is also 
natural from the perspective of fuzzy spaces \cite{Sasakura:2011ma,Sasakura:2005js}. 
In the interpretation, a configuration $X_{abc}$ of the tensor model is assumed to 
correspond to a fuzzy space defined by an algebra of the functions $f_a$ on it,
\[
\label{eq:defoffuzzy}
f_a \cdot f_b=X_{abc} f_c.
\]
Here $X_{abc}$ plays the role of the structure constants of the function algebra.
Since the essential properties of the functions do not change under the common rescaling
$f_a\rightarrow \gamma f_a \hbox{ for all }f_a$, imposing the gauge symmetry \eq{eq:gammax}
is a natural requirement.

The above discussions imply the necessity of adding a new constraint
${\cal D}=0$ with
\[
{\cal D}=\frac{\sigma}{3} X_{abc} Y_{abc},
\]
which generates a scaling transformation,
\[
\label{eq:transdx}
&\{{\cal D},X_{abc}\}=X_{abc},\\
\label{eq:transdy}
&\{ {\cal D}, Y_{abc} \}= -Y_{abc}.
\]
The newly introduced ${\cal D}$ forms a closed algebra 
with ${\cal J}_{[ab]}$ and ${\cal H}_a$ as
\[
\label{eq:algdh}
&\{{\cal D}, H(T) \}=H(T),\\
\label{eq:algdj}
&\{{\cal D}, J(V) \}=0.
\]
The algebraic on-shell closure of 
\eq{eq:alghh}, \eq{eq:algjh}, \eq{eq:algjj}, \eq{eq:algdh}, \eq{eq:algdj} 
on the constraint subspace ${\cal H}_a={\cal J}_{[ab]}={\cal D}=0$ 
implies that a new canonical tensor model can 
consistently be defined as a constraint system 
${\cal H}_a={\cal J}_{[ab]}={\cal D}=0$\footnote{In fact, it seems possible
to consider a shifted constraint ${\cal D}-d=0$ with a non-zero real parameter $d$. 
This ambiguity may be avoided by embedding the algebra into a larger one,
which has ${\cal D}$ as a result of Poisson brackets between constraints.
This possibility is left for future study.}.

\section{The configuration space and classical fixed points}
\label{sec:classical}
The total hamiltonian of the new system is given by 
\[
H^{new}_{tot}={\cal N}_a {\cal H}_a + {\cal N} {\cal D}+{\cal N}_{[ab]} {\cal J}_{[ab]},
\]
where ${\cal N}$ is a new variable.
Then the equation of motion is given by 
\[
\label{eq:newceqm}
\frac{d X_{abc}}{dt}=\{X_{abc},H^{new}_{tot}\}\approx 
-\sigma {\cal N}_d 
\left(X_{d(ae)}X_{ebc}+X_{d(be)}X_{eca}+X_{d(ce)}X_{eab}
\right)
-\sigma {\cal N} X_{abc}
+{\cal N}_{[de]}(\cdots),
\]
where ${\cal H}_a$ is taken to be \eq{eq:hycde}.

Since the trivial configuration, $^\forall X_{abc}=0$, is a fixed point of the classical equation
of motion \eq{eq:newceqm}, one cannot get to it with a finite time 
starting from another configuration. Therefore one can consistently decouple the 
trivial point from the rest of the configuration space. 
By using ${\cal D}$, which generates \eq{eq:transdx},
an arbitrary configuration in the rest space can be gauge fixed as
\[
\label{eq:xx1}
X_{abc}X_{abc}^*=1.
\]
Thus the configuration space of the new model can be 
represented by the intersection of the compact space \eq{eq:xx1} and some other
gauge-fixing conditions.
In such a compact space, classical solutions will in general 
asymptotically approach either fixed points or cyclic orbits,
but will not have the extreme behaviors as the previous model explained in 
Section~\ref{sec:new}. 

It is not difficult to give a general example for fixed points of the classical equation of motion
\eq{eq:newceqm}.
Suppose that there exists an index value $0$, which satisfies
\[
X_{0ab}=x_0 \delta_{ab}
\]
with a real parameter $x_0$. Suppose also 
a gauge which takes ${\cal N}_a=n_0 \delta_{0a}$ with a
real parameter $n_0$ and ${\cal N}_{[ab]}=0$.
Then the equation of motion \eq{eq:newceqm} becomes
\[
\frac{d X_{abc}}{dt}&= 
-\sigma n_0
\left(X_{0(ae)}X_{ebc}+X_{0(be)}X_{eca}+X_{0(ce)}X_{eab}
\right)
-\sigma {\cal N} X_{abc},\nonumber \\
&=-\sigma(3 n_0 x_0+{\cal N}) X_{abc}. 
\]
Then a fixed point solution can be obtained by $x_0=-{\cal N}/3 n_0$. 
One can further set ${}^\forall Y_{abc}=0$ for the equation of motion of $Y_{abc}$
and the constraints to be satisfied.

The above setup for fixed point solutions naturally appears 
for configurations with group symmetries.
To see this,
consider a configuration $\bar X_{abc}$ which is invariant under 
a group $L$ embedded in $O(N)$ 
as\footnote{For concrete examples, $\bar X_{abc}$ can be taken to be 
C-G coefficients among
various representations of groups such as the 3$j$-symbol of SO(3).} 
\[
\label{eq:xinv}
l_a{}^d l_b{}^e l_c{}^f \bar X_{def}=\bar X_{abc}, \ \ ^\forall l\in L \subset O(N). 
\]  
Here the representation of $L$ on $\bar X_{abc}$ is assumed to be reducible
to a number of irreducible representations by the $O(N)$ transformation
and contain uniquely a one-dimensional trivial representation.
Then 
\[
\label{eq:x0ab}
\bar X_{0ab}=x_{R(a)} \delta_{ab},
\] 
where $0$ denotes the index value in the trivial representation, and $x_{R(a)}$ are 
real parameters which can depend on each irreducible 
representation $R(a)$ to which the index value $a$ belongs.
On such a symmetric configuration, 
one can in principle take a gauge which is consistent 
with the group symmetry. This requires ${\cal N}_a=n_0 \delta_{0a}$,
and that ${\cal N}_{[ab]}$ take a gauge in which ${\cal N}_{[ab]}{\cal J}_{[ab]}$
generates the infinitesimal transformation of the group symmetry 
\eq{eq:xinv}\footnote{If the group symmetry does not have infinitesimal transformations,
such as in case of a discrete symmetry,
${\cal N}_{[ab]}$ are taken to vanish.}. 
Then, since $\{X_{abc}, {\cal N}_{[de]}{\cal J}_{[de]} \}|_{X=\bar X} \approx 0$ 
because of  \eq{eq:xinv},
the situation becomes the same as the 
previous paragraph, and 
\[
\label{eq:valxra}
x_{R(a)}=-{\cal N}/3 n_0
\]
is a fixed point of the classical equation of motion.

It is noteworthy that the above solution satisfies a simple usual property of 
a space, when it is interpreted as a fuzzy space defined by \eq{eq:defoffuzzy}.
From \eq{eq:x0ab} and \eq{eq:valxra}, one obtains,
after proper rescaling of $f_a$ with ${\cal D}$\footnote{
This is equivalent to take a gauge ${\cal N}=-3 n_0$.}, 
\[
f_0 \cdot f_a=f_a\ \hbox{ for all }f_a.
\]
This implies that there exists a constant function $f_0$ on the space. 
This is actually non-trivial, since a fuzzy space defined by \eq{eq:defoffuzzy} does not 
necessarily have such
a constant function for general $X_{abc}$.

\section{The minimal tensor model}
\label{sec:uniqueness}
Because of a vast number of degrees of freedom of tensor models, 
it should be useful to think of a minimal model.
This is the tensor model with a real symmetric rank-three
tensor,
\[
\label{eq:symmetric}
&X_{abc}^*=X_{abc}, \\
&X_{abc}=X_{bca}=X_{cab}=X_{bac}=X_{acb}=X_{cba}.
\]
In the canonical formalism, $X=M,P$.
In this section, I will discuss the uniqueness of the local Hamiltonian 
constraint of the canonical real symmetric rank-three tensor model
with the new constraint ${\cal D}=0$.

The (two-fold) uniqueness \eq{eq:hycde}, \eq{eq:hyced} of the local Hamiltonian constraint
shown in the previous paper \cite{Sasakura:2012fb} is only
for the canonical rank-three tensor model with the Hermiticity condition \eq{eq:hermiticity}.
In fact, if the dynamical variables are the totally symmetric real tensors 
\eq{eq:symmetric} and the new constraint ${\cal D}=0$ is {\it not} introduced, 
the most general form of the local Hamiltonian constraint
under the physically reasonable assumptions of the previous paper
can be shown to have a one-parameter ambiguity as 
\[
\label{eq:hsym}
{\cal H}^{sym,no{\cal D}}_a=
X_{abc}X_{bde}Y_{cde}+\lambda Y_{abb},
\]
where $\lambda$ is an arbitrary real constant. 
This can easily be shown 
by applying the former part of the previous paper \cite{Sasakura:2012fb} to
this case, and checking the on-shell algebraic closure\footnote{ 
The most difficult issue in the previous paper  
was how to incorporate the 
complications originating with the change of orders of the indices of $M_{abc}$
and $P_{abc}$, 
since it
generates quite a large number of distinct terms which potentially 
compose a local Hamiltonian constraint. 
This issue was treated in the latter part of the previous paper, after 
the former part of the analysis ignoring the orders. On the other hand, 
in the present case,
$M_{abc}$ and $P_{abc}$ are symmetric and therefore the former part is enough.
The conclusion of the former part is that the diagrams $G^4$ and $G^1$ 
are allowed, which correspond to
the two terms in \eq{eq:hsym}, respectively.}. 

From \eq{eq:transdx} and \eq{eq:transdy}, one can see that 
the two terms in \eq{eq:hsym} are transformed differently by ${\cal D}$.
Therefore, $\lambda=0$ is required 
for the algebraic closure of the constraints, ${\cal H}_a,\ {\cal J}_{ab},
\ {\cal D}$.
Thus 
\[
{\cal H}^{sym}_a=X_{abc}X_{bde}Y_{cde}
\]
is the unique local Hamiltonian constraint 
for the real symmetric rank-three tensor model
with the constraints, ${\cal H}_a={\cal J}_{ab}={\cal D}=0$.

\section{Summary and future prospects}
\label{sec:summary}
The canonical rank-three tensor model proposed in the previous paper has the
bad feature that the solutions to the classical equation of motion have extreme behaviors.
There exist no other fixed points other than the trivial one, and
the classical solutions either diverge or asymptotically vanish in time evolution.  
These extreme behaviors would  
become major obstacles in future study such as of
obtaining stationary spaces and quantizing the model.  
   
To improve the previous model, 
this paper has proposed a new canonical rank-three tensor model by adding 
a scaling constraint.
This constraint is a natural expectation from the interpretation that
the rank-three tensor model describes dynamics of fuzzy spaces.
The new constraint makes the configuration space compact, and the classical solutions 
asymptotically approach either fixed points or cyclic orbits in general.
It is shown that configurations with group symmetries 
provides a general class of such fixed points.
These fixed points would represent stationary fuzzy spaces in physical 
interpretation of the model.

With the scaling constraint, it is also shown that 
the local Hamiltonian constraint is unique in the minimal case, namely,
the canonical real symmetric rank-three tensor model.
This is in contrast with that, without the scaling constraint, the local Hamiltonian 
has one parameter ambiguity.
The new canonical symmetric rank-three tensor model will provide 
the simplest setup for future study.

It would obviously be interesting to study the large $N$ dynamics of the canonical tensor models.
Since the local Hamiltonians have rather simple polynomial forms, the corresponding Lagrangians and hence 
the Feynman rules will become involved. This in turn would potentially make the large $N$ behaviors of the 
canonical tensor models significantly different from those of the unsymmetric tensor models 
\cite{Gurau:2011xp,Bonzom:2012wa,Gurau:2013cbh,Dartois:2013he}.  
Or an alternative way of study would be to carry out perturbative expansions around fixed points
discussed in Section~\ref{sec:classical}. In this case, the fixed points would provide backgrounds, 
and the situation would rather have similarity to the formalism of 
the tensor group field theories \cite{BenGeloun:2011rc,BenGeloun:2012pu,Geloun:2012bz,Carrozza:2012uv}.

\section*{Acknowledgement}
The author would like to thank L.~Freidel for discussions on tensor models and 
various other topics of quantum gravity, which have much influenced the contents 
of the present paper, during his stay in YITP as the visiting professor of Kyoto University. 
The author would also like to thank L.~Freidel, V.~Bonzom, and J.B.~Geloun 
for invitation, hospitality and stimulating discussions on tensor models 
during his stay in Perimeter Institute.


\end{document}